# Self-assembly of InAs nanostructures on the sidewalls of GaAs nanowires directed by a Bi surfactant

*Ryan B. Lewis\*, Pierre Corfdir, Jesús Herranz, Hanno Küpers, Uwe Jahn, Oliver Brandt and Lutz Geelhaar*

Paul-Drude-Institut für Festkörperelektronik, Hausvogteiplatz 5–7, 10117 Berlin, Germany

*Email: lewis@pdi-berlin.de

Surface energies play a dominant role in the self-assembly of three dimensional (3D) nanostructures. In this letter, we show that using surfactants to modify surface energies can provide a means to externally control nanostructure self-assembly, enabling the synthesis of novel hierarchical nanostructures. We explore Bi as a surfactant in the growth of InAs on the $\{1\bar{1}0\}$ sidewall facets of GaAs nanowires. The presence of surface Bi induces the formation of InAs 3D islands by a process resembling the Stranski−Krastanov mechanism, which does not occur in the absence of Bi on these surfaces. The InAs 3D islands nucleate at the corners of the $\{1\bar{1}0\}$ facets above a critical shell thickness and then elongate along $\langle 110 \rangle$ directions in the plane of the nanowire sidewalls. Exploiting this growth mechanism, we realize a series of novel hierarchical nanostructures, ranging from InAs quantum dots on single $\{1\bar{1}0\}$ nanowire facets to zig-zag shaped nanorings completely encircling nanowire cores. Photoluminescence spectroscopy and cathodoluminescence spectral line scans reveal that small surfactant-induced InAs 3D islands behave as optically active quantum dots. This work illustrates how surfactants can provide an unprecedented level of external control over nanostructure self-assembly.

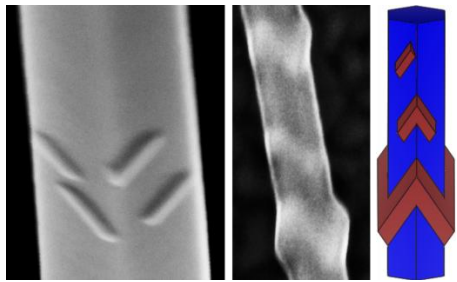

The bottom-up self-assembly of semiconductor nanostructures is directed by energy minimization. As a result, many aspects of nanostructure self-assembly are intrinsic and therefore difficult to control externally. For example, nanostructures such as nanowires (NWs) and quantum dots (QDs) spontaneously form low-energy facets during their synthesis. In the case of GaAs NWs grown by the Ga-assisted vapor-liquid-solid (VLS) mode, the sidewall facets are of {1$\bar{1}$0} orientation.[1] This poses challenges for the realization of advanced hierarchical structures, such as QDs embedded within NWs,[2,3] since two-dimensional (2D) layer growth is always favored on GaAs{110} surfaces and three-dimensional (3D) islands do not form by the Stranski–Krastanov (SK) mechanism.[2,4–6] The SK growth of InAs 3D islands on these surfaces has been observed after covering the facets with a thin AlAs layer, but, this is undesirable for most applications.[2,7] In contrast, GaAs NWs synthesized by Au-catalyzed growth typically exhibit {112} sidewall facets and the SK mechanism does occur on these surfaces,[6,8] however, the presence of Au can be detrimental to NW optoelectronic properties.[9]

We recently reported that the presence of surface Bi can induce the self-assembly of InAs 3D islands directly on planar GaAs(110) surfaces.[10] The Bi surfactant was shown to modify the surface energies, reducing the energetic cost of 3D island formation. Furthermore, in contrast to more common surface-segregating elements like Sb and Te, which have been shown to reduce adatom diffusion and consequently inhibit the formation of 3D islands,[11,12] Bi has been found to increase adatom diffusion.[13] This makes Bi of particular interest for III-V NW synthesis, where adatom diffusion is an essential factor. However, Bi and surfactants in general remain almost completely unexplored within the context of NW growth, despite the exceedingly important role of surface effects due to the high surface to volume ratio of NWs. Initial investigations have shown that the presence of Bi during Au-catalyzed GaAs NW synthesis can alter the NW crystal structure from wurtzite to zincblende, possibly a result of the surfactant effect of Bi.[14] The growth of GaAs/Ga(As,Bi) core-shell NWs has also recently been reported.[15] Such structures offer the potential to realize GaAs-based optoelectronics operating at longer wavelengths than those possible with conventional alloying elements such as In and Sb, owing to the large reduction of the GaAs band gap obtained by Bi alloying.[16]

In this work, we show that surfactant Bi can induce the formation of InAs 3D nanostructures directly on the {1$\bar{1}$0} sidewalls of GaAs NWs. The surface Bi provokes the nucleation of 3D InAs islands at the corners of the {1$\bar{1}$0} facets, while growth in the absence of Bi results in a 2D shell. With continued InAs deposition, the 3D islands elongate along ⟨110⟩ directions in the plane of the NW sidewalls, which we exploit to realize a series of novel nanostructures ranging from InAs 3D islands to zig-zag shaped nanorings. The small 3D islands behave as optically active QDs, demonstrating their perspective for quantum optics embedded in GaAs NWs operating at telecommunications wavelengths. This work illustrates that surfactants can open the door to new possibilities for hierarchical nanostructure self-assembly.

We first explore the effect of a Bi flux during the deposition of thin InAs shells around GaAs NWs, on the shell morphology. Figure 1 presents scanning electron microscopy (SEM) images of single NWs of about 60 nm diameter after deposition of 2.1 monolayers (MLs) of InAs on the {1$\bar{1}$0} NW sidewalls under various Bi beam equivalent pressures (BEPs). We note that 2.1 MLs on {1$\bar{1}$0} corresponds to 1.5 MLs on {100}. In the absence of Bi, a thin shell segment of about 200 nm in length is visible in the lower part of the image [Figure 1(a)]. Investigation of multiple NWs from this sample reveal similar features, indicating that the InAs deposition produces a discontinuous shell, consistent with previous reports.[2] With increasing Bi BEP, the aspect ratio and thickness of the features on the NW sidewalls increase. At an intermediate Bi BEP of 5×10$^{-7}$ mbar [Figure 1(b)], the sidewall features have shrunk to less than 100 nm in length and still show flat {1$\bar{1}$0} tops. For deposition under the highest Bi BEP of 2×10$^{-6}$ mbar [Figure 1(c)], the NW facets are decorated with 3D islands of typical base diameters of 20–40 nm and heights of 5–9 nm. The increasing aspect ratio of the sidewall features with increasing Bi flux is consistent with our recent study of InAs deposition on planar GaAs(110) using Bi as a surfactant. In this previous study, density functional theory calculations showed that surface Bi reduces the driving force for InAs wetting by strongly reducing the surface energy of the bare GaAs(110) surface.[10] This surface energy reduction allows the system to favor strain relaxation through the formation of 3D islands. For the As-rich growth conditions used here, we do not expect substantial Bi incorporation.[17,18] We also note that for InAs QDs grown on planar GaAs(100), the use of Bi as a surfactant was shown to significantly increase the photoluminescence intensity of the QDs as well as to redshift their energy, which is of interest for realizing QDs emitting at telecommunication wavelengths.[19]

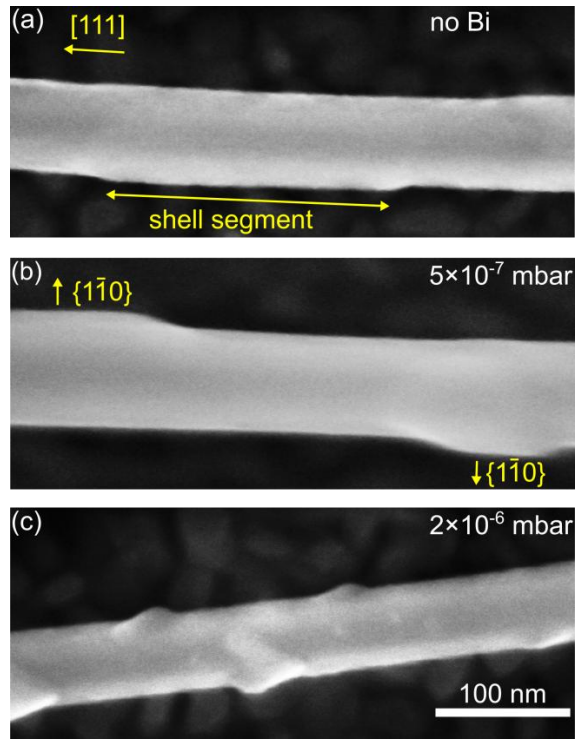

**Figure 1**. SEM images of dispersed GaAs NWs of about 60 nm diameter after deposition of 2.1 MLs of InAs on the sidewalls under Bi BEPs indicated in the figure. (a) In the absence of Bi, very low aspect ratio InAs shell segments are observed. (b)–(c) With increasing Bi BEP, the aspect ratio and thickness of the segments increases, eventually resulting in 3D islands. The viewing direction is normal to the NW axis.

In order to better visualize the geometry of the 3D islands on the NW sidewall, InAs islands were deposited on thicker NWs of about 170 nm diameter under a Bi BEP of $2\times10^{-6}$ mbar. These thick NWs were obtained by depositing a GaAs shell before the InAs deposition. Figure 2 shows SEM images of two NWs from this sample, where the viewing direction is inclined by 20° from the NW axis and such that the edge between two {1 $\bar{1}$ 0} sidewalls is observed in the center of each NW. We find that the 3D islands are always in contact with an edge separating two {1 $\bar{1}$ 0} facets, indicating that the 3D islands nucleate at the edges of the {1 $\bar{1}$ 0} facets. This is likely a result of the favorable strain relaxation resulting from the convex character of the edges.[20] We note that in a previous study of strain relaxation in (In,Ga)As shells grown on GaAs NWs of similar dimensions, the nucleation of larger (In,Ga)As mounds was also observed to occur at the edges of the sidewall facets.[21] In Figure 2(a), the 3D islands are highly asymmetric, being elongated in the ⟨110⟩ direction in the plane of the {1 $\bar{1}$ 0} sidewalls. While the 3D islands in Figure 2(b) also show sides running parallel to ⟨110⟩ directions, they are smaller and more symmetric (note that the imaging direction results in a compression along NW axis of about 3 times). We speculate that these 3D islands nucleated at a later stage of the growth and thus have had less time to elongate. We note that InAs islands grown on planar GaAs(110) in the presence of Bi were also found to be elongated in the [1 $\bar{1}$ 0] direction.[10] This elongation may be the result of asymmetric adatom diffusion on the {1 $\bar{1}$ 0} surface, or a consequence of the elastic anisotropy, as InAs is stiffer along the in-plane ⟨001⟩ direction than along the orthogonal ⟨110⟩ direction.

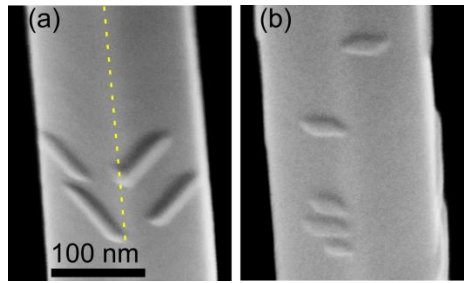

**Figure 2**. SEM images of as-grown NWs of about 170 nm diameter after deposition of 2.1 MLs of InAs under a Bi BEP of $2\times10^{-6}$ mbar. The 3D islands nucleate at the edges of the {1 $\bar{1}$ 0} sidewall facets (indicated in (a) by a dotted line) and elongate along ⟨110⟩ directions in the plane of the sidewalls (35.3° from the [111] NW axis). The viewing direction is 20° from the NW axis.

The sidewall morphologies of dispersed NWs of about 60 nm diameter with varying nominal InAs thicknesses deposited under a Bi BEP of $2\times10^{-6}$ mbar are shown in Figure 3. For 0.8 MLs of InAs [Figure 3(a)], the NW sidewalls show no sign of 3D islands. This indicates the existence of a critical thickness for 3D island formation, as is the case for the SK mechanism. After 1.4 ML of InAs deposition [Figure 3(b)], 3D islands become visible on the NW sidewalls. For this sample, individual islands are found to mostly occupy only single sidewall facets. As the InAs thickness is further increased to 2.1 MLs, individual nanostructures elongate and begin occupying multiple facets of the NWs [Figure 3(c)]. Since the elongation direction is always along ⟨110⟩ directions in the plane of the nanowire sidewalls, when a 3D island crosses to an adjacent {1$\bar{1}$0} facet the axial component of the elongation direction reverses. This peculiar growth process results in V-shaped 3D islands when a single nanostructure occupies two adjacent {1$\bar{1}$0} facets. Finally, for the largest deposition thickness of 3.5 MLs [Figure 3(d)], many structures completely encircle the NW core, producing well defined zig-zag shaped nanostructures. Two presentations of an atomic force microscopy (AFM) topograph of a single NW after 2.1 MLs of InAs deposition are shown in Figures 3(e) and 3(f), revealing an assortment of nanostructures covering both single and multiple facets. These features have typical heights of 5−9 nm, as demonstrated by the profiles shown in the inset of Figure 3(e). The InAs nanostructures are found to exhibit a pair of {111} facets on each {1$\bar{1}$0} NW sidewall facet. Figure 3(g) illustrates the family of nanostructures composed of {111} facets. Nanostructures occupying adjacent {1$\bar{1}$0} NW facets share one {111} facet. For zig-zag shaped nanorings encircling the GaAs core, the nanostructure is completely defined by 6 {111} facets and the NW core. These findings illustrate that Bi not only provokes the self-assembly of InAs 3D islands on the {1$\bar{1}$0} sidewalls of GaAs NWs, but the well defined elongation of these islands allows for an array of novel hierarchical nanostructures to be realized. This growth mode could be used to fabricate quantum rings in NWs.

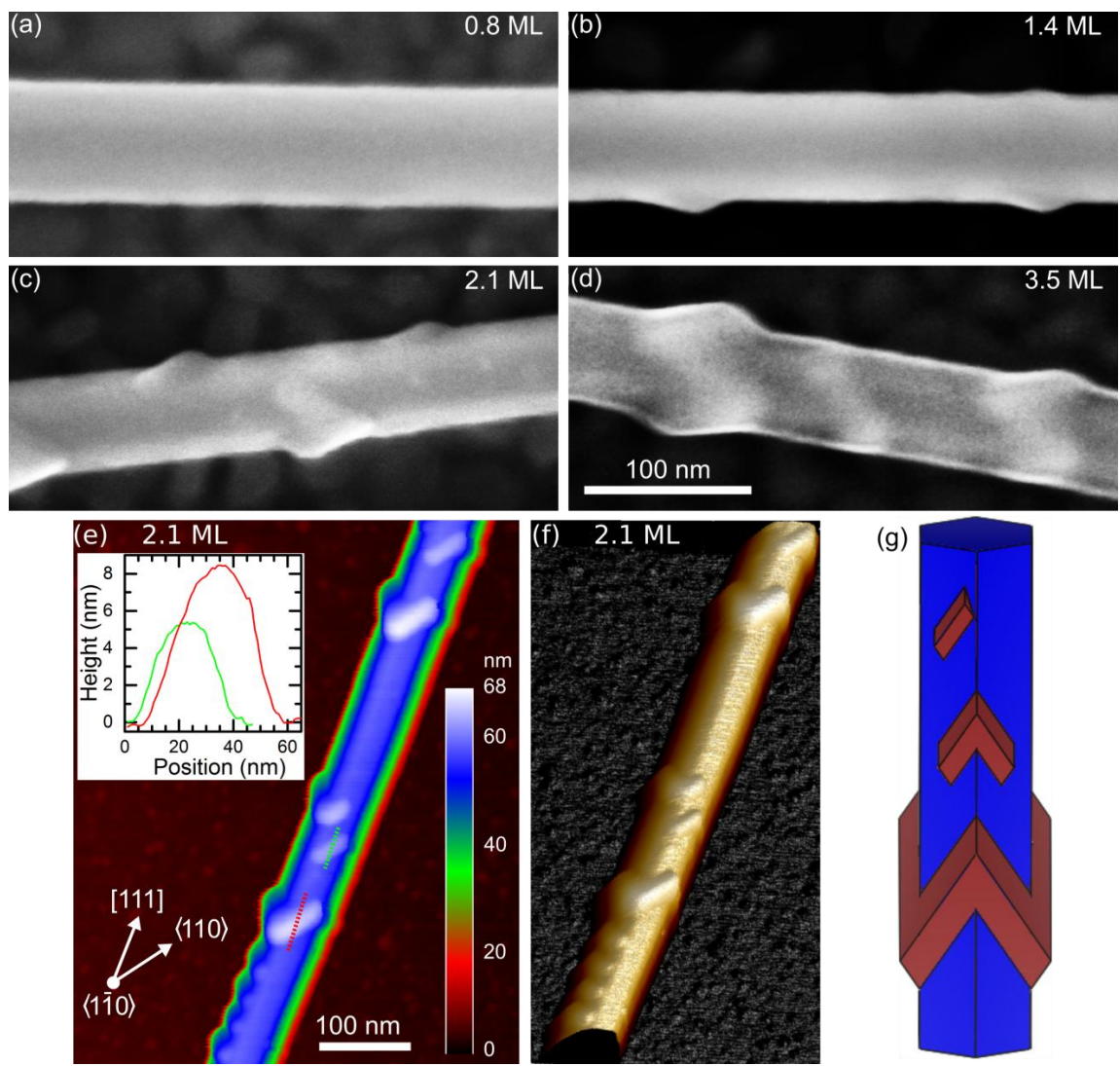

**Figure 3**. (a)−(d) SEM images of dispersed NWs of about 60 nm diameter after depositing InAs with nominal thicknesses indicated in the figure under a Bi BEP of $2\times10^{-6}$ mbar. The viewing direction is normal to the NW axis. (a) No indication of 3D growth is observed after 0.8 MLs of InAs. (b) After 1.4 MLs of InAs, 3D objects appear on the NW sidewalls. (c) With 2.1 MLs of deposition, the 3D nanostructures are observed to occupy multiple sidewall facets, forming V-shaped nanostructures as they cross the facet edges. (d) At 3.5 MLs InAs deposition, zig-zag nanorings completely encircling the NW core are observed. (e)−(f) AFM topographs of the same NW after 2.1 ML of InAs showing an assortment of 3D nanostructures composed of {111} facets and covering both single and multiple facets. The height profiles from the dotted lines in (e) are shown in the inset. (g) Illustration of the observed nanostructures composed of {111} facets.

We have shown that the presence of a Bi flux can provoke the formation of InAs 3D islands directly on the {1$\bar{1}$0} sidewalls of GaAs NWs, while the absence of Bi results in a discontinuous InAs shell. To explore the impact of these morphological changes on the NW optical properties, as-grown NWs containing 1.4 MLs of InAs deposited with and without the presence of Bi and capped with GaAs/AlAs/GaAs shells were investigated by photoluminescence (PL) and cathodoluminescence (CL) spectroscopies. Figure 4(a) presents PL spectra from these samples. Given the low NW density of about 0.3 μm$^{-2}$ and the small excitation spot size of about 1 μm$^2$, we expect that the PL spectra originate from individual NWs. The PL spectrum for the sample grown without Bi is dominated by a band at about 1.33 eV, which we associate with emission from the thin InAs shell. This spectrum exhibits some narrow features with a full width at half maximum (FWHM) of at least 1 meV, which we attribute to local fluctuations in shell width and possibly In-content (due to intermixing), as well as the fact that the shell is polytypic.[22] Two spectra are presented for the sample grown with Bi, which are dominated by a band centered at 1.44 eV. On the lower energy side of this band, additional sharp transitions are detected. As shown in the inset of Fig. 4(a), the FWHM for the transitions in the 1.3–1.4 eV energy range is about 120 μeV, which corresponds to the resolution of our optical system. We attribute these changes in PL properties to the fact that when grown using Bi surfactant, the InAs shell consists of a thin wetting layer covered by 3D islands, giving rise to the broad transition at 1.44 eV and to the narrow lines in the 1.3–1.4 eV energy range, respectively. The linewidth of the 1.3–1.4 eV transitions is one order of magnitude lower than what has been reported for excitons localized by interface roughness or stacking faults in (In,Ga)As quantum well shells in nanowires,[22,23] suggesting that the microscopic origin of the centers localizing excitons in samples grown with Bi is different. Furthermore, transitions of comparable linewidths have been observed in the same energy range in the PL spectra of InAs 3D islands grown on planar GaAs(110) substrates using Bi.[10] In that study, the recombination originated from neutral excitons confined in small InAs islands of 1−2 nm height and about 10 nm diameter. We therefore conclude that the transitions observed in the 1.3−1.4 eV energy range in Fig. 4(a) result from recombination of excitons in InAs islands of similar size present on the NW sidewalls (smaller than those observable with SEM). Finally, we note that luminescence from the GaAs NW core could not be detected, as carriers generated in the core are captured by the InAs shell, which is beneficial for increasing the carrier capture by the InAs QDs. This finding is in contrast to previously reported InAs 3D islands realized on GaAs{1$\bar{1}$0} NW sidewalls covered with a thin AlAs layer.[2]

To investigate the spatial distribution of the emission features observed in PL, we carried out CL spectral line scans on single as-grown NWs. Figure 4(b) presents an exemplary CL line scan along the length of a single NW grown with Bi, where the electron beam was inclined by 45° from the NW axis. Along the entire length of the NW, the CL spectrum is dominated by the broad emission band centered at 1.44 eV, attributed to exciton recombination in the InAs wetting layer. The sharp emission features in the energy range of 1.3−1.4 eV are spatially localized along the NW length, consistent with recombination associated with the InAs 3D islands. We note that these emission features are elongated along the NW axis, which we attribute to diffusion of carriers excited by the electron beam to the QDs. Accounting for the finite width of the generation volume in the CL experiments,[24] as well as the 45° angle between the NW axis and the electron beam, we deduce an axial carrier diffusion length of about 400 nm from the spatial widths of these emission features along the length of the NW. This value is comparable to previously reported diffusion lengths in GaAs/(Al,Ga)As axial NW heterostructures passivated with an (Al,Ga)As shell.[25] In contrast, for GaAs NWs with (Al,Ga)As shell QDs it was reported that carrier diffusion to the QDs was negligible, as carriers excited in the (Al,Ga)As shell are efficiently captured by the GaAs core.[26] This is in opposition to the present case, where the InAs QDs can capture carriers from the core, explaining the lack of GaAs core luminescence. These results indicate that Bi can induce the formation of optically active InAs QDs directly on the {1$\bar{1}$0} sidewalls of VLS grown GaAs NWs.

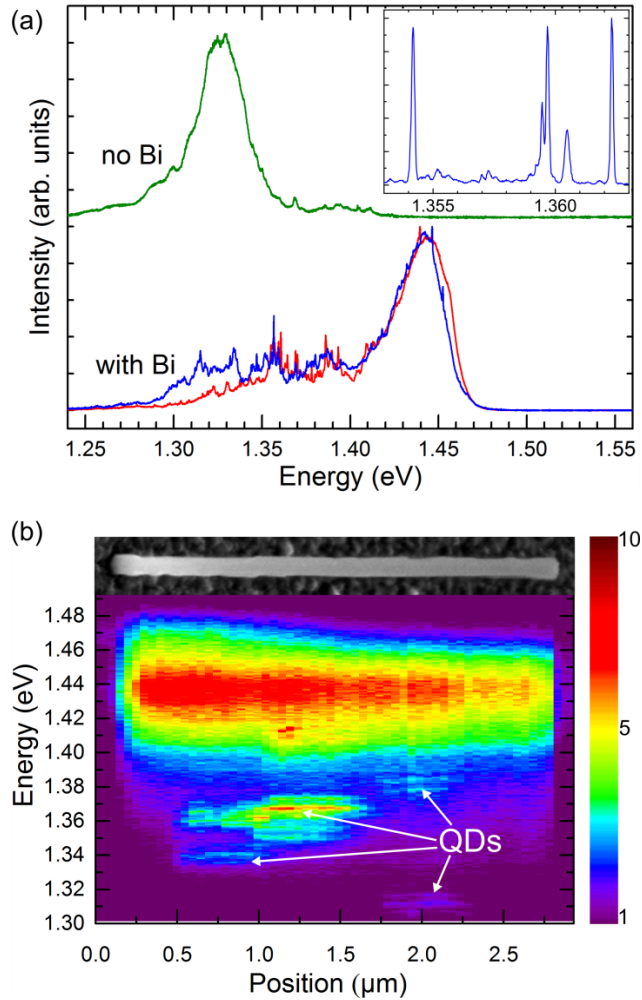

**Figure 4.** (a) PL spectra ($\lambda_E$ = 632.8 nm, 9 K) taken from as-grown capped NW samples containing 1.4 MLs of InAs grown with and without the presence of a Bi flux. Scans from two regions of the sample grown with Bi show a series of narrow transitions in the 1.3–1.4 eV energy

range, and the wetting layer emission is blue-shifted compared to the sample grown without Bi. The inset shows transitions with FWHMs of about 120 μeV (resolution limited) from the sample grown with Bi ($\lambda_E$ = 650 nm, 5 K). (b) A 2D plot of CL spectra (linear intensity scale shown on right) recorded at 8 K along the axis of a single NW grown with Bi (SEM image displayed above). The plot shows wetting layer emission at about 1.44 eV that extends along the length of the NW, as well as spatially localized emission lines between 1.3 and 1.4 eV. The electron beam was inclined by 45° from the NW axis.

In summary, we have demonstrated the surfactant-induced self-assembly of InAs 3D nanostructures directly on the {1Ī0} sidewalls of GaAs nanowires, surfaces on which 3D growth is otherwise inhibited. The InAs 3D islands form by a Stranski–Krastanov-like mechanism, nucleating at the edges of the {1Ī0} nanowire facets and then elongating along ⟨110⟩ directions in the plane of the nanowire sidewalls. We exploit this controlled growth process to realize nanostructures ranging from 3D islands on single {1Ī0} facets to novel zig-zag nanoring structures encircling the nanowire core. The small 3D islands behave as optically active quantum dots, demonstrating perspectives for optoelectronic devices operating at telecommunications wavelengths, as well as optomechanical applications. Given the crucial role of surface effects for nanostructure self-assembly, this work illustrates that surfactants can provide an unprecedented level of external control over the synthesis of these structures.

## AUTHOR INFORMATION

**Corresponding Author**

*Email: lewis@pdi-berlin.de## METHODS

Samples were grown by molecular beam epitaxy on Si(111) wafers covered by native oxide. The $As_2$ flux was provided by a valved cracker and Ga (2 cells), In and Bi were provided by conventional effusion cells. Substrate temperatures were measured by a pyrometer calibrated to the oxide desorption temperature of GaAs(100). After annealing the untreated wafers at about 650°C for 10 min, the substrate temperature was reduced to 630°C, after which Ga was deposited on the substrate for 30 s at 0.37 nm/s GaAs equivalent growth rate followed by a 60 s flux interruption to establish Ga droplets on the surface.[21,27] GaAs NW cores with density, diameter and length on the order of 0.3 μm$^{-2}$, 60 nm and 7 μm, respectively, were subsequently grown by Ga-assisted VLS growth with respective Ga and $As_2$ deposition rates of 0.085 nm/s and 0.71 nm/s. Following the NW core growth, the Ga droplets atop the NWs were consumed by exposure to $As_2$ and the substrate temperature was reduced to 420°C while maintaining an $As_2$ flux of 1.1 nm/s and 0.17 Hz substrate rotation. A Bi flux was initiated 30 s before the InAs deposition and maintained throughout. During InAs deposition, the In flux on the NW sidewalls (averaged over the substrate rotation) was 0.021 ML/s. Samples for optical characterization were capped with GaAs/AlAs/GaAs with respective shell thicknesses 5 nm/10 nm/5 nm, with shell growth rates of 0.012 nm/s under an $As_2$ flux of 1.1 nm/s. Uncapped samples were cooled at 2°C/s to below 350°C under the $As_2$ flux.

All PL and CL experiments were carried out on as-grown nanowires. For the PL experiments realized at 9 K, the samples were mounted on the coldfinger of a continuous-flow He cryostat, and a HeNe laser emitting at 632.8 nm was used for excitation. The laser beam was focused using a 0.7 N.A. microscope objective, and the PL signal was dispersed and detected with a monochromator equipped with a 600 lines/mm grating and a liquid $N_2$-cooled charge-coupled device, respectively. For the experiments carried out at 4.2 K, the samples were kept at liquid He temperature in a confocal setup. A laser diode emitting at 650 nm was used for excitation, and the laser beam was focused using a 0.82 N.A. microscope objective. The PL signal was collected using the same objective and coupled to a single mode fiber, whose core acts as a confocal hole. The signal was then dispersed using an 1800 lines/mm grating and detected with a liquid $N_2$-cooled charge-coupled device. The CL measurements were performed in a field-emission scanning electron microscope equipped with a Gatan monoCL4 system and a He-cooled stage. CL spectra along the NW axis (CL line scans) were recorded at a sample temperature of 8 K using a charge-coupled device detector. The electron acceleration voltage and beam current were 5 kV and 0.3 nA, respectively. The angle between the electron beam and the NW axis was 45°.

ACKNOWLEDGMENTS

R.B.L. acknowledges funding from the Alexander von Humboldt Foundation. P.C acknowledges funding from the Fonds National Suisse de la Recherche Scientifique through Project No. 161032. This work was partially funded by Deutsche Forschungsgemeinschaft under grant Ge2224/2. The authors are grateful to M. Höricke and C. Stemmler for MBE maintenance, M. Ramsteiner and G. Paris for help with the PL setup, and E. Zallo for a critical reading of the manuscript.